\documentclass{article}

\usepackage[preprint, nonatbib]{neurips_2022}

\usepackage[utf8]{inputenc} 
\usepackage[T1]{fontenc}    
\usepackage{hyperref}       
\usepackage{url}            
\usepackage{booktabs}       
\usepackage{amsfonts}       
\usepackage{nicefrac}       
\usepackage{microtype}      
\usepackage{lipsum}
\usepackage{graphicx}
\usepackage{subfigure}
\graphicspath{ {./images/} }
\usepackage{xspace} 
\usepackage{amsmath}
\usepackage{amssymb}

\begin{document}

\title{Coherent noise suppression via a self-supervised blind-trace deep learning scheme}
\author{
 Sixiu Liu, Claire Birnie, Tariq Alkhalifah \\
  King Abdullah University of Science and Technology, \\
  \texttt{\{sixiu.liu, claire.birnie, tariq.alkhalifah\}@kaust.edu.sa} \\
}
\date{} 
\maketitle

\begin{abstract}
Coherent noise regularly plagues seismic recordings, causing artefacts and uncertainties in products derived from down-the-line processing and imaging tasks. The outstanding capabilities of deep learning in denoising of natural and medical images have recently spur a number of applications of neural networks in the context of seismic data denoising. A limitation of the majority of such methods is that the deep learning procedure is supervised and requires clean (noise-free) data as a target for training the network. Blindspot networks were recently proposed to overcome this requirement, allowing training to be performed directly on the noisy field data as a powerful suppressor of random noise.  A careful adaptation of the blind-spot methodology allows for an extension to coherent noise suppression. In this work, we expand the methodology of blind-spot networks to create a blind-trace network that successfully removes trace-wise coherent noise. Through an extensive synthetic analysis, we illustrate the denoising procedure's robustness to varying noise levels, as well as varying numbers of noisy traces within shot gathers. It is shown that the network can accurately learn to suppress the noise when up to 60\% of the original traces are noisy. Furthermore, the proposed procedure is implemented on the Stratton 3D field dataset and is shown to restore the previously corrupted direct arrivals. Our adaptation of the blind-spot network for self-supervised, trace-wise noise suppression could lead to other use-cases such as the suppression of coherent noise arising from wellsite activity, passing vessels or nearby industrial activity. 
\end{abstract}


\section{Introduction}
In seismic surveys, field data are often contaminated by coherent noise that exhibit
coherency in time and/or offset, which makes seismic processing and imaging difficult. Often, the existence of such noise may mask the useful reflection signal \cite{claerbout1985ground,henley2003coherent}, which can lead to unreliable subsequent inversion results \cite{yilmaz2001seismic},
and thus should be removed early in seismic processing to improve data quality. Coherent noise,
such as trace-wise noise caused by poorly coupled receivers, is one such cause of coherent
noise in seismic records.

Typically noise attenuation approaches are subdivided into four general categories. The first category includes prediction-based methods that utilize both the predictable property of useful signals and the unpredictable property of noise to design prediction filters. For example, the polynomial fitting (PF) method has been used to suppress noise by estimating local linear coherent components in seismic data \cite{lu2006local,liu2011seismic}. The second category of suppression procedures include sparse transform-based approaches that separate signals and noise in a transformed domain. Signals and noise are sparsely represented, separated by thresholding coefficients in the transformed domain and transformed back to the original domain. These methods include the Fourier transform \cite{yilmaz2001seismic}, the radial trace (RT) transform \cite{claerbout1985ground,henley2003coherent,henley2003more} and the wavelet transform (WT) \cite{deighan1997ground,goudarzi2012seismic}, among others. Similarly, mathematical morphological filtering and 
median filtering attenuate linear noise based on the difference in the shape of seismic waves in the transformed coordinate system \cite{huang2017mathematical} and the energy difference between
noise pixels and the median value of the neighborhood \cite{weihua2000coherent,dong2018blending,chen2020deblending}, respectively. In some cases, differences between useful signals and noise in a transformed domain or a coordinate system are too small to be distinguished. Therefore, the applications of the second category are limited. The third category, decomposition-based approaches, decompose noisy data into different components and then recreate the useful signals based on the principal components. For example, singular value decomposition (SVD) \cite{kendall2005svd,cary2009ground,porsani2009ground} and empirical mode decomposition (EMD) \cite{bekara2009random,gomez2016simple} have both been illustrated as successful suppressors of coherent noise. The last category, rank-reduction-based approaches, assume the seismic data to be of low-rank after some data rearrangement. For example, multichannel singular spectrum analysis (MSSA) was employed to denoise a 3D cross-spread data set that was severely contaminated with ground roll \cite{chiu2013coherent}. 

Recently, deep learning has been widely used in the field of geophysics and has made breakthroughs in seismic denoising. Among these are Convolutional Neural Networks (CNNs)-based models for random noise removal, such as standard CNNs \cite{mandelli2019interpolation,yuqing2019random,wang2019applying}, CNNs with residual learning \cite{zhang2018noise,jin2018seismic,wang2019residual}, CNNs with transfer learning \cite{yu2019deep} and denoising convolutional neural network (DnCNN) \cite{liu2018random,li2021deep}. CNNs-based models have also proven successful at coherent noise attenuation within seismic data. For example, standard CNNs has been utilized for the likes of deblending \cite{baardman2019classification,sun2020convolutional} and linear noise removal \cite{ma2018deep}. The DnCNN architecture has also been employed for the suppression of linear noise within seismic data \cite{zheng2020improved,yuan2021attenuation}. 

Conventional deep learning noise suppression procedures typically take noise-free seismic sections
as labels for training the neural network in a supervised way to attenuate noise in seismic data. However, for field data, noise-free records are not available. Different from supervised learning techniques that require labeled data for training, AutoEncoders (AEs) have the ability to automatically learn and store features of the unlabeled, noisy data in an unsupervised learning manner. Working similarly to traditional decomposition-based procedures, the denoising ability of AEs are exploited for both random \cite{chen2019improving,song2020seismic} and coherent noise suppression \cite{hamidi2020autoencoder}. However, the learned features by AEs are not sufficient to fully represent seismic data \cite{zhang2019seismic}. The latent space in AEs is typically small, making the AEs unable to reconstruct all the seismic features accurately. This motivates the investigation of other self-supervised procedures, which only use the noisy data as the training samples. Recently, a self-supervised denoising method, Noise2Void (N2V) was proved to effectively attenuate random noise in seismic data in a self-supervised manner by training directly on the data to be denoised \cite{birnie2021self,birnie2021potential}. This is achieved by masking the center pixel (blind spot) of the receptive field of the network \cite{krull2019noise2void}. N2V can be considered as a prediction-based approach due to it's ability to accurately recreate the coherent signal and it's inability to predict the noise component due to the noise's random nature. A similar masking strategy was employed by \cite{meng2021self} to train a CNN over a J-invariant function for seismic random noise suppression in a self-supervised manner. Within the field of microscopy, \cite{broaddus2020removing} proposed an extension of N2V, termed StructN2V, where the blind-spot is extended beyond a single pixel to create a mask that conceals the coherency of noise. This masking scheme was successfully shown to suppress linear fluorescence noise. An initial investigation into the use of such networks for processing seismic data was proposed by \cite{liu2022self}.

In this work, we first introduce our adaptation of the StructN2V approach to effectively remove trace-wise correlated noise in seismic data using only noisy images as both the input and target for the blind-trace networks. The performance of the proposed algorithm is evaluated on synthetic and field data examples. We further investigate the proposed algorithm from the perspective of differing noise levels and varying levels of noisy-traces within the seismic data.
\section{Theory}
In conventional CNN procedures for denoising, a pixel value within an image is predicted based on the values of pixels within a neighboring square (receptive field) centralised on the predicted pixel. Hereon in, we shall refer to this central pixel as the active pixel. As the receptive field contains the active pixel, the active pixel's value in the input is utilised to predict the active pixel's value in the target image. \cite{krull2019noise2void} proposed the use of a blind-spot scheme, to be applied on an active pixel, to avoid its value influencing the network's prediction of the pixel's value. Under the assumption of random noise and coherent signal, employing such a scheme prohibits the network from learning the noise component of the pixel's value and therefore, only the signal component can be predicted.

A limitation of the blind-spot implementation of \cite{krull2019noise2void} is that it requires noise to be independent between pixels. To overcome this, \cite{broaddus2020removing} proposed the extension of the blind-spot to create a ‘blind-patch’, which we will refer to as noise mask throughout this study. The mask is designed  prior to training to hide the coherent noise information in neighbouring pixels from one specific direction, which is calculated by cross correlation of the pure noise.  However, accurately designing the noise mask in relation to the noise is challenging. As signals are closely correlated between pixels and vital for predicting signal components of pixels values, improper selection of the mask's orientation has the potential to conceal the signals as opposed to the desired noise concealment. The size of the mask also needs  careful consideration. A larger mask size will potentially hide the signals, information that is required to predict the signal value of an active pixel. A smaller mask may not be big enough to hide all the noise information, and therefore might cause noise leakage into the value of a predicted pixel. In our case, the noise in seismic data we are targeting is trace-wise dependent along the whole trace. As such, a vertical mask across the whole trace of seismic shot records is used. 

\begin{figure}
    \centering
    \includegraphics[width=1\textwidth,height=0.5\textwidth]{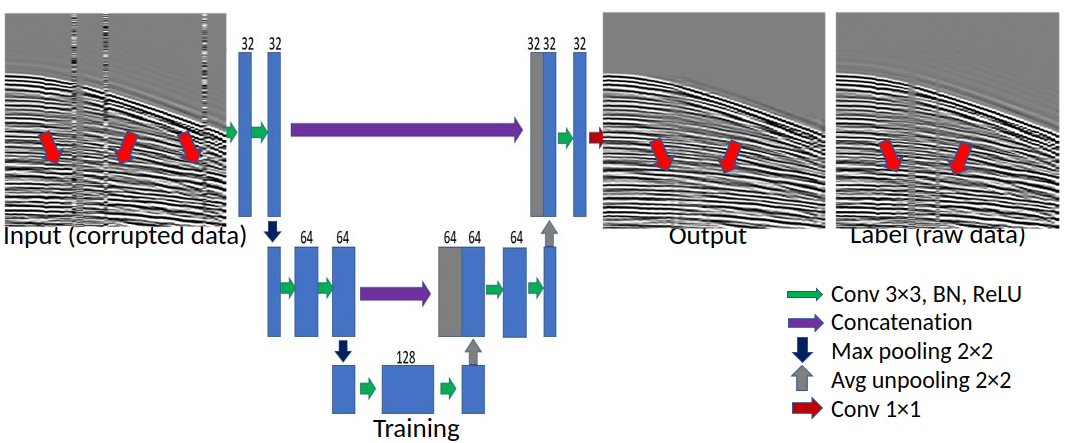}
    \caption{Schematic diagram of the training workflow using StructN2V to suppress coherent noise. Red arrows in input, output and label illustrate three corrupted traces, three denoised traces and the original three noisy traces, respectively. }
    \label{fig:theory}
\end{figure}

The noise mask is implemented as a pre-processing step on the raw, noisy, shot gathers prior to being passed to the neural network. In this masking scheme, first some pixels are selected as the active pixels for training. These active pixels may or may not fall on traces contaminated with trace-wise noise. Then, all pixels belonging to the same trace location of the active pixels are replaced (i.e., masked) by random values drawn from a uniform distribution, as illustrated in Figure \ref{fig:theory} by red arrows. The purpose of the noise mask is to make the noise within the masked locations unpredictable prior to being passed into the network. After training, a pixel’s signal value can still be predicted from its neighbouring pixels (outside the mask) due to its spatial predictability along the offset, but the noise component cannot be predicted due due to its temporal unpredictability along the time axis.

As illustrated in Figure \ref{fig:theory}, a standard U-Net architecture is utilized 
following other blind-spot denoising applications \cite{broaddus2020removing,birnie2021potential,liu2022self}. The corrupted versions of raw noisy shot gathers created through the mask scheme and the original noisy shot gathers are regarded as the inputs and the labels (targets) to the network during training, respectively. In the process of training, the network can learn to recreate the seismic arrivals due to their spatial properties however cannot recreate the coherent noise. Therefore, whilst the label is the noisy data, the output should be the denoised result, as illustrated in Figure \ref{fig:theory}. Instead of computing all the 
pixels for loss as done in classical deep learning methods, the loss function here is computed only on the corrupted pixels. As part of our experiments, we have considered the loss on both the active pixels and across the full masked trace.  It should be noted that the denoising loss is not desired to be zero, which is the case in traditional supervised learning methods, otherwise the identity of the noisy image is learned and the network will only output the original noisy image. The loss $L$ is computed as follows, 

\begin{equation}
\begin{split}
    \mathcal{L} &=\sum_{j}\sum_{i}\mathcal{L}_{d}\left(f\left(\tilde{\boldsymbol{x}}_{\Omega }^{ji};\boldsymbol{\theta}\right),\boldsymbol{x}^{ji}\right) ,
\end{split}
\end{equation}

where $i$ and $j$ are the number of active pixels within an image and the number of the training samples, respectively. The trace that is randomized by the blind-trace operation is represented by $\Omega$. The value of the $i$th active pixel value within the $j$th training label (i.e., raw patch) and the value of the $i$th active pixel value within the $j$th training input (i.e., corrupted patch) are represented by $\boldsymbol{x}^{ji}$ and $\tilde{\boldsymbol{x}}_{\Omega }^{ji}$, respectively. The trained network architecture and network parameters are denoted by $f$ and $\theta$, respectively. The loss function applied to either only active pixels (field case) or the full masked trace (synthetic case), and the total loss normalised by Mean Absolute Error (MAE) are represented by $\mathcal{L}_{d}$ and $\mathcal{L}$, respectively. After a specified number of iterations in the optimization process, the network should hopefully have learnt how to recreate the desired signal without recreating the targeted noise in the noisy data.

\section{Numerical Examples}
We evaluate the approach on two different datasets: synthetic seismic data with different levels and different numbers of noisy traces, and a field seismic dataset. In order to quantitatively compare the quality of the denoised results with corresponding noise-free data, two performance metrics are used: the  Image Peak Signal-to-Noise Ratio (PSNR) and the Structural Similarity Index Measure (SSIM). PSNR is a classic seismic metric whilst SSIM is the preferred metric from the deep learning community for denoising applications.

PSNR is expressed as
\begin{equation}
P S N R=10 \log _{10} \frac{\max\|D\|_{2}^{2}}{\|D-d\|_{2}^{2}} ,
\end{equation}
where $D$ and $d$ stand for the clean data and the compared data (either noisy or denoised). The unit of PSNR is decibel (dB). The greater the PSNR value, the better the denoised result.

SSIM measures the similarity between two images ($D$ and $d$) and it is expressed as
\begin{equation}
SSIM(D,d)=\frac{\left(2\mu_D\mu_d+c_1\right)\left(2\sigma_{Dd}+c_2\right)}
{\left(\mu_D^2+\mu_d^2+c_1\right)\left(\sigma_D^2+\sigma_d^2+c_2\right)} ,
\end{equation}
where the average of $D$ and the average of $d$ are denoted by $\mu_D$ and $\mu_d$, respectively. The variance of $D$ and the variance of $d$ are represented by $\sigma_D^2$ and $\sigma_d^2$, respectively. The covariance of $D$ and $d$ is $\sigma_{Dd}$. Two variables to stabilize the division with weak denominator are $c_1=(k_1L)^2$ and $c_2=(k_2L)^2$. The dynamic range of the pixel value is $L$. By default, $k_1=0.01$ and $k_2=0.03$. The structural similarity ranges from 0 to 1. The greater the SSIM value, the better the denoised result. When the two images are identical, the value of SSIM is equal to 1.
\subsection{Synthetic example}
\begin{figure}[htp]
    \centering
    \includegraphics[width=1\textwidth,height=0.4\textwidth]{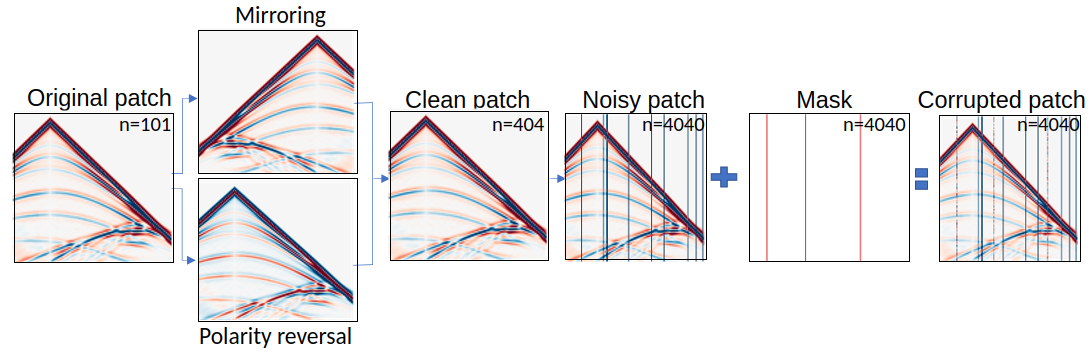}
    \caption{Data preparation workflow.}
    \label{fig:Datapreparation}
\end{figure}

The proposed Self-Supervised Learning (SSL) procedure is first validated on synthetic seismic records constructed via acoustic, finite-difference modelling of the SEG Hess VTI model, shown as the “original patch” in Figure \ref{fig:Datapreparation}. Figure \ref{fig:Datapreparation} shows how the training data are prepared using data augmentation procedures to increase the training data size, where $n$ represents the number of the training patches generated at the different stages of the pipeline. Initially, 101 shots are generated and are then increased four-fold by the data augmentation techniques of mirroring and polarity reversal. In order to generate noisy patches, ten different noise realisations are created for each shot in the last stage by randomly replacing ten traces with a random value between [-0.8, 0.8] to represent the raw, noisy data used as the target (label) for the deep learning network. The blind-trace denoising scheme is then employed on each of the noisy patches, in which the masked traces, shown by red vertical lines in the mask, will be corrupted by random values from a uniform distribution over [-0.8, 0.8] to generate the input data, referred to corrupted patch. The corrupted noise is preferred to have a similar magnitude as the “recorded” noise to avoid changing the raw, noisy image too much while employing the masking scheme. This corruption scheme is implemented per patch and per epoch, therefore changing the input to the network on each training iteration. A ReLU activation function and batch normalisation are used in the convolutional layers, and a linear activation function is used on the output layer. The training parameters were optimised via hyper-parameter tuning following a grid-search procedure, with the optimal parameters detailed in Table \ref{tab:para}.

\begin{table}[htp]
\centering
\caption{Training parameters for the two data sets}
\label{tab:params}
\scalebox{0.9}{%
\begin{tabular}{ |p{3cm}|p{2cm}|p{2cm}|  }
\hline
   & Synthetic &Field \\
\hline
Patch size & 512x256 &56x56 \\
Train:Val & 3200:840   & 7000:1140 \\
 \# Active Traces  &3 & 3 \\
Noise level    &[-0.8, 0.8] & [-1.5, 1.5] \\
UNet depth  & 4 & 3 \\
Batch size & 64 & 16   \\
\# Initial filters  & 32 & 32 \\
Kernel  & 3 & 3 \\
LR   & 0.004 & 0.004 \\
Loss   & MAE & MAE \\
 \# Epochs   & 128 & 72 \\
Loss coverage   & trace & active \\
\hline
\end{tabular}}
\label{tab:para}
\end{table}

\begin{figure}[htp]
    \centering
    \includegraphics[width=1\textwidth,height=0.5\textwidth]{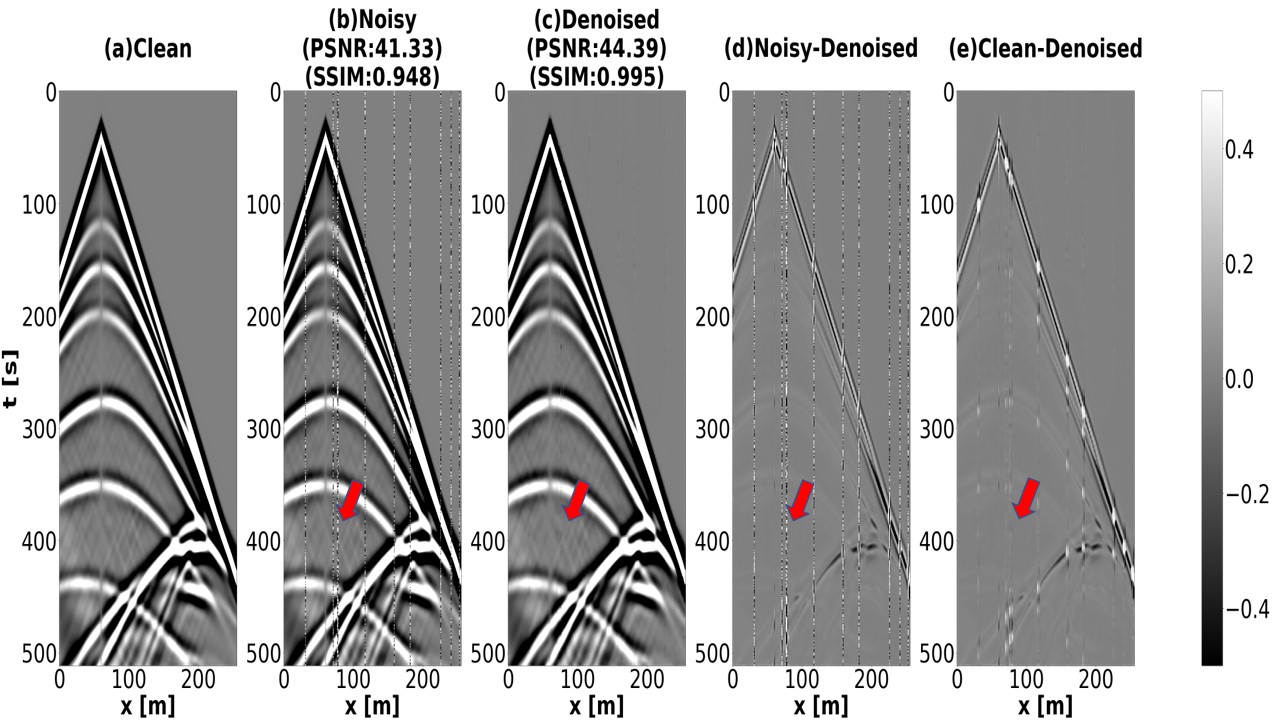}
    \caption{Trained SSL model applied to a synthetic dataset with coherent noise.  (a) The noise-free synthetic, (b) the noisy synthetic given as input to the model,and  (c)  the  result  of  the  SSL  denoising  procedure.   (d)  and  (e)  are  the  differences between the noisy and denoised datasets and between the noise-free and denoised datasets, respectively. Red arrow in (b), (c) and (d) illustrate the noisy traces, the denoised traces and the removed noise, respectively.}
    \label{fig:synthetic}
\end{figure}

The results from the denoising scheme applied to the Hess VTI synthetic seismic dataset trained with the blind-trace procedure are shown in Figure \ref{fig:synthetic}, alongside the observed PSNR and SSIM. It can be seen that some of the signal in the raw, noisy data (Figure \ref{fig:synthetic}(b)) is hidden  by the strong trace-wise noise. Compared to the noisy data shown in Figure \ref{fig:synthetic}(b) with a PSNR of 41.33 dB and a SSIM of 0.948, the denoised result shown in Figure \ref{fig:synthetic}(c) has a higher PSNR of 44.39 dB and a higher SSIM of 0.995. The denoised result illustrates almost no remaining  coherent noise. Figure \ref{fig:synthetic}(d) and (e) highlight the noise removed by the method and the difference between the noise-free and denoised datasets, respectively, both of which show that the proposed method can accurately suppress the trace-wise noise. However, these figures also show the existence of a slight difference in the amplitudes of the first arrival, indicating slight signal leakage for the high amplitude first arrival.

\paragraph{Effect of noise level}
To further evaluate the denoising performance, we apply different levels of trace-wise noise contamination for synthetic seismic shot gathers. After obtaining the 404 clean created for each shot by randomly replacing 10 traces among the total 256 traces with random noise level ranging from 0.1 to 0.9, with step of 0.1, to represent the 4040 raw, noisy data with different noise levels as the target (label) for the deep learning network. For comparitive purposes, the value ranges for the first arrival and the reflections are [-4.4, 8.2] and [-0.6, 1], respectively. For post-training analysis, 100 additional noisy shot gathers are generated, again by randomly replacing 10 traces with random values over the pre-described range. All the training parameters are as detailed in Table \ref{tab:para} and are the same for the nine different cases, except for the noise levels to corrupt traces. The PSNR and SSIM are both used to evaluate the denoising performance and are averaged across 100 test images. 

\begin{figure}[htp]
    \centering
    \subfigure[\centering PSNR Change ]{{\includegraphics[width=0.46\textwidth,height=0.4\textwidth]{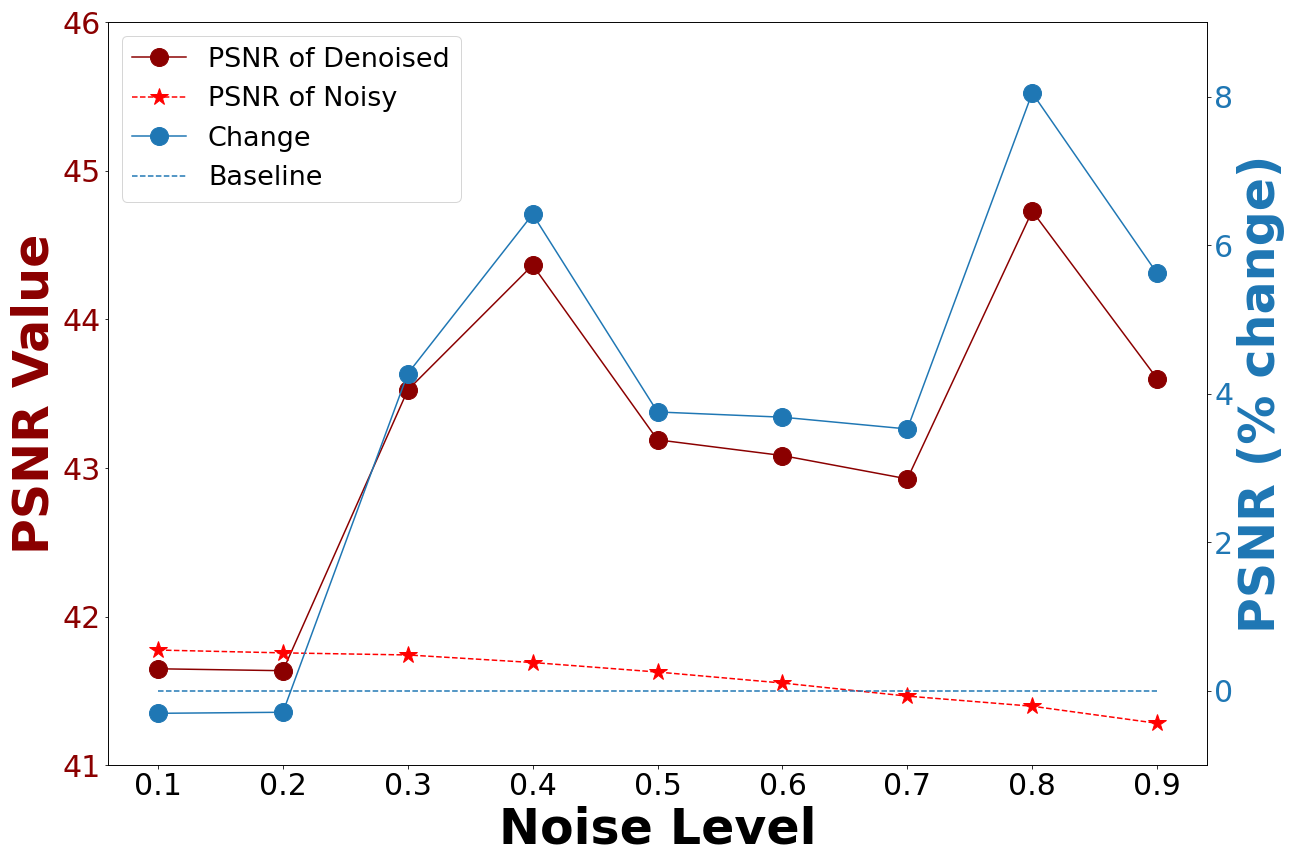} }}%
    \qquad
    \subfigure[\centering SSIM Change ]{{\includegraphics[width=0.46\textwidth,height=0.4\textwidth]{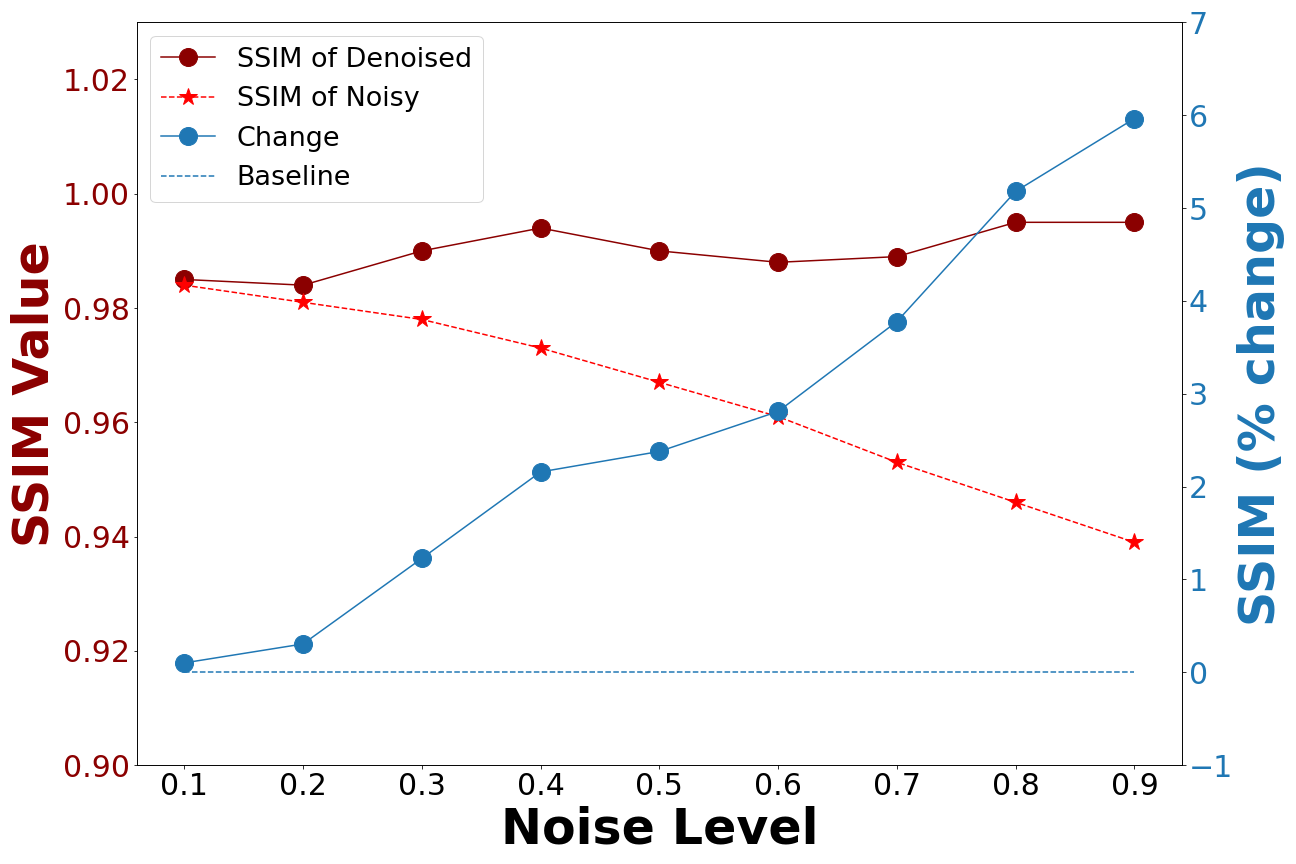} }}%
    \caption{PSNR value and PSNR change over noise levels (a), and SSIM value and SSIM change over noise levels (b). The noisy data has a consistent standard deviation of 0.1 and 0.002 for PSNR and SSIM, respectively.}%
    \label{fig:Levelexample}%
\end{figure}

Figure \ref{fig:Levelexample} illustrates the relationship between data quality (PSNR and SSIM) values and percent changes over noise contamination level (the amplitude values of noisy traces in a patch). As shown in Figure \ref{fig:Levelexample}(a), as the noise level of noisy traces increases, the denoised PSNR value and change (between the noisy input data and the denoised results) both increase with a drop between 0.4-0.8. All the changes for noise levels larger than 0.2 are above zero, which means the denoised result has a higher PSNR than the noisy image, indicating a higher tolerance with higher noise level. The fluctuation in PSNR is possibly due to the random elements introduced during the training - corruption values and locations. In Figure \ref{fig:Levelexample}(b), the SSIM change increases gradually while the denoised SSIM value increases slowly as the contamination level increases. Similar to the PSNR,  the denoised result has a higher SSIM value than the noisy image as the changes are always above zero. Both the PSNR and the SSIM of the noisy data decrease gradually due to the introduction of higher noise contamination level. The proposed procedure can be considered as a stable and powerful denoiser for different levels of trace-wise noise.

\paragraph{Effect on number of noisy traces}
To conclude the synthetic analysis, we investigate the effect of including varying numbers of noisy traces within the ‘raw’ shot gathers. After obtaining the 404 clean patches in Figure \ref{fig:Datapreparation}, 10 different noise realisations are generated
for each shot by randomly replacing 10-90\% of the traces among the total 256 traces with a random value between [-0.8, 0.8]. These represent the 4040 raw, noisy data samples as the target (label) for the deep learning network. As before, 100 new test images are generated following the same procedure as the training data generation. All the training parameters are as detailed in Table \ref{tab:para} and are the same for the nine different cases. Again, the PSNR and SSIM are both used to evaluate the denoising performance and are averaged across the 100 test images. 

\begin{figure}[htp]
    \centering
    \subfigure[\centering PSNR Change ]{{\includegraphics[width=0.46\textwidth,height=0.4\textwidth]{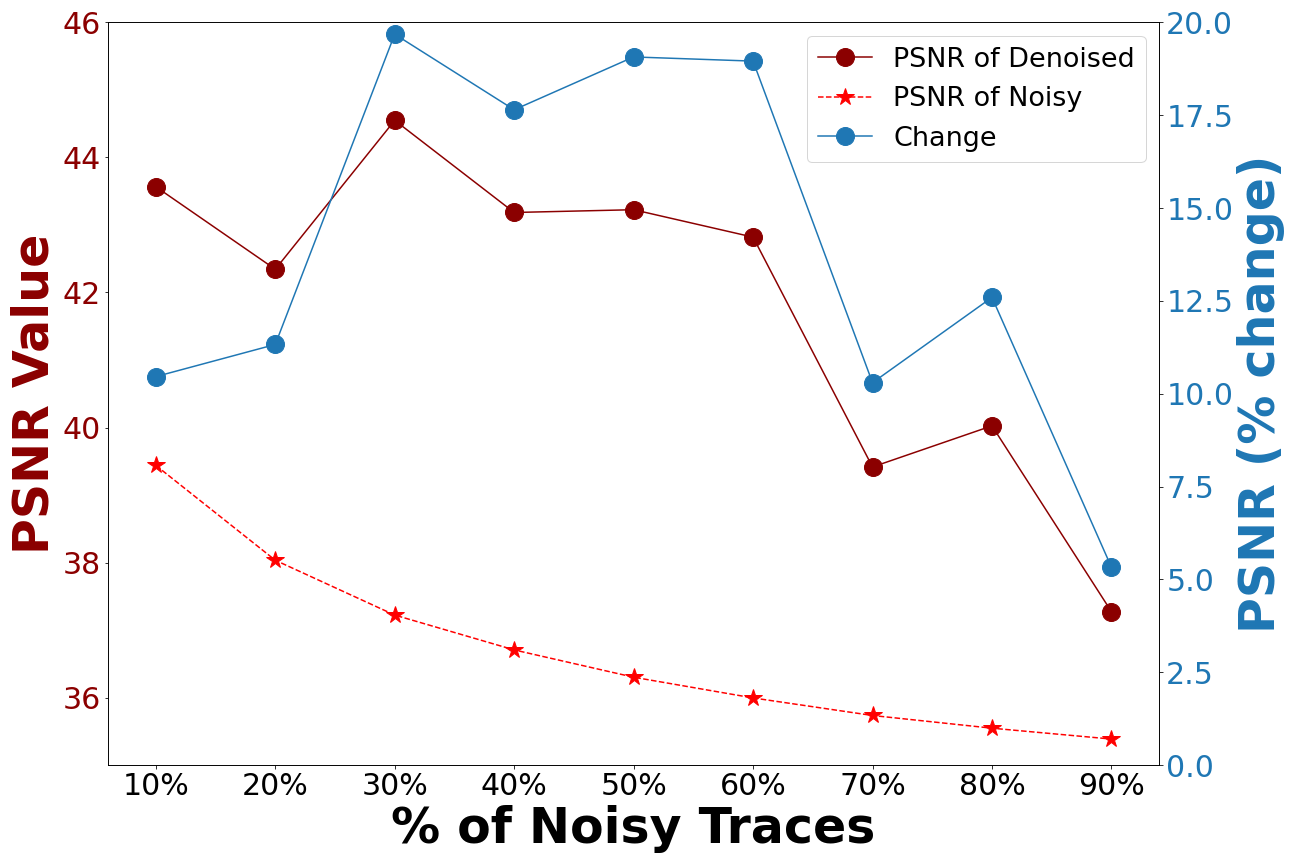} }}%
    \qquad
    \subfigure[\centering SSIM Change ]{{\includegraphics[width=0.46\textwidth,height=0.4\textwidth]{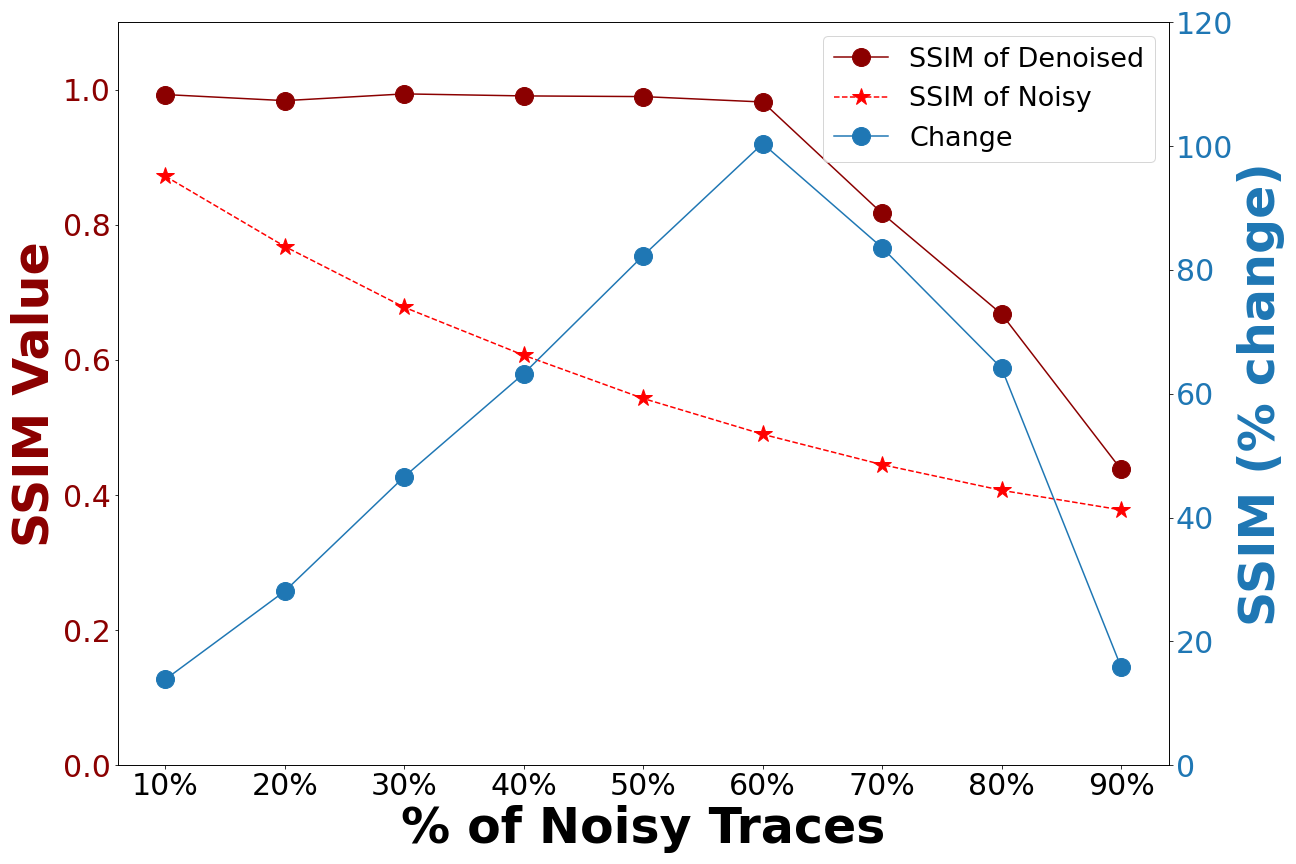} }}%
    \caption{PSNR value and PSNR change over numbers of noisy traces (a), and SSIM value and SSIM change over numbers of noisy traces (b). The noisy data has a consistent standard deviation of 0.1 and 0.02 for PSNR and SSIM, respectively.}%
    \label{fig:percentexample}%
\end{figure}

Figure \ref{fig:percentexample} illustrates the relationship between the data quality (PSNR and SSIM) values and percent changes over the percent of noisy traces. As shown in Figure \ref{fig:percentexample}(a), as the percentage of noisy traces increases, both of the denoised values and changes of PSNR increase before 30\% and then drop, due to the introduction of more trace-wise noise. In Figure \ref{fig:percentexample}(b), the denoised SSIM value stays just below $1.0$ and the SSIM percent change increases almost linearly before 60\%. At around 60\%, both the denoised SSIM value and the percent change start to drop, highlighting that the denoising network is no longer working under its optimal conditions. Both the PSNR and the SSIM of the noisy data decrease gradually due to the introduction of higher number of noisy traces. Both plots in Figure \ref{fig:percentexample} show a positive change in SSIM and PSNR, indicating the network can still successfully suppress a portion of the noise when up to 90\% of the raw shot is contaminted by noise.  Therefore, the method can denoise trace-wise noise contamination percentage of 90\% traces in seismic data. The method is shown to be optimal where less than 60\% of the raw traces are noisy.
\subsection{Field data application}
\begin{figure}[htp]
    \centering
    \includegraphics[width=1\textwidth,height=0.5\textwidth]{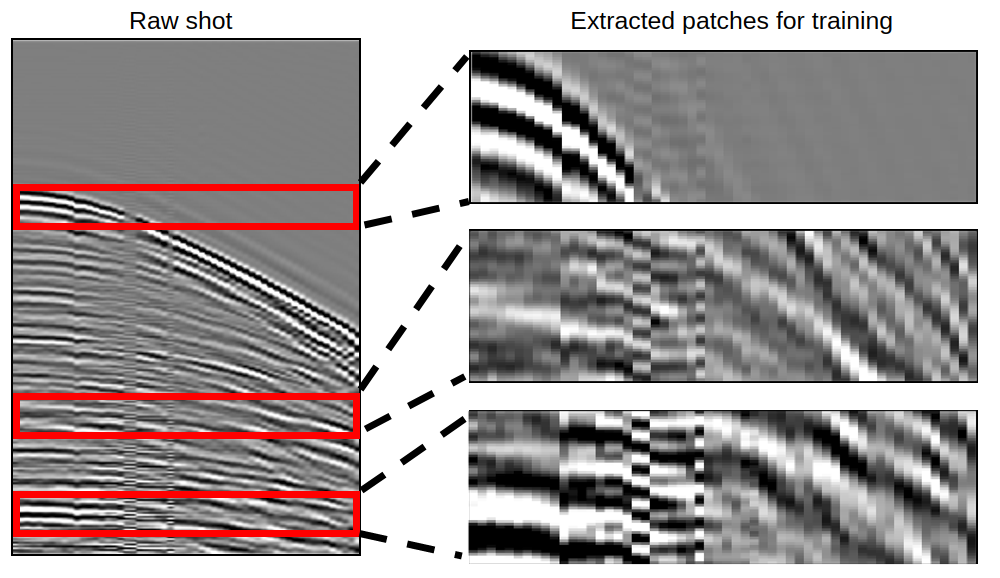}
    \caption{Schematic of the data patching workflow for Stratton 3D land dataset. On the left is the full shot gather whilst on the right are three extracted patches.}
    \label{fig:field patches}%
\end{figure}

Shot gathers from the Stratton land 3D dataset from South Texas are used to verify the practical effectiveness of the proposed method. For this field data example, a data patching technique is applied on the original noisy images of size $1024\times56$ to generate 8140 noisy patches of size $56\times56$, as shown in Figure \ref{fig:field patches}. In other words, the data are patched into time windows containing 56 samples, 112 ms, covering the full receiver array. Patching is utilised to ensure adequate data is available for the network training. We take 7000 noisy patches with $56\times56$ pixels as our training dataset and 1140 noisy patches of the same size as the test dataset. 

\begin{figure}[htp]
    \centering
    \includegraphics[width=1\textwidth,height=1\textwidth]{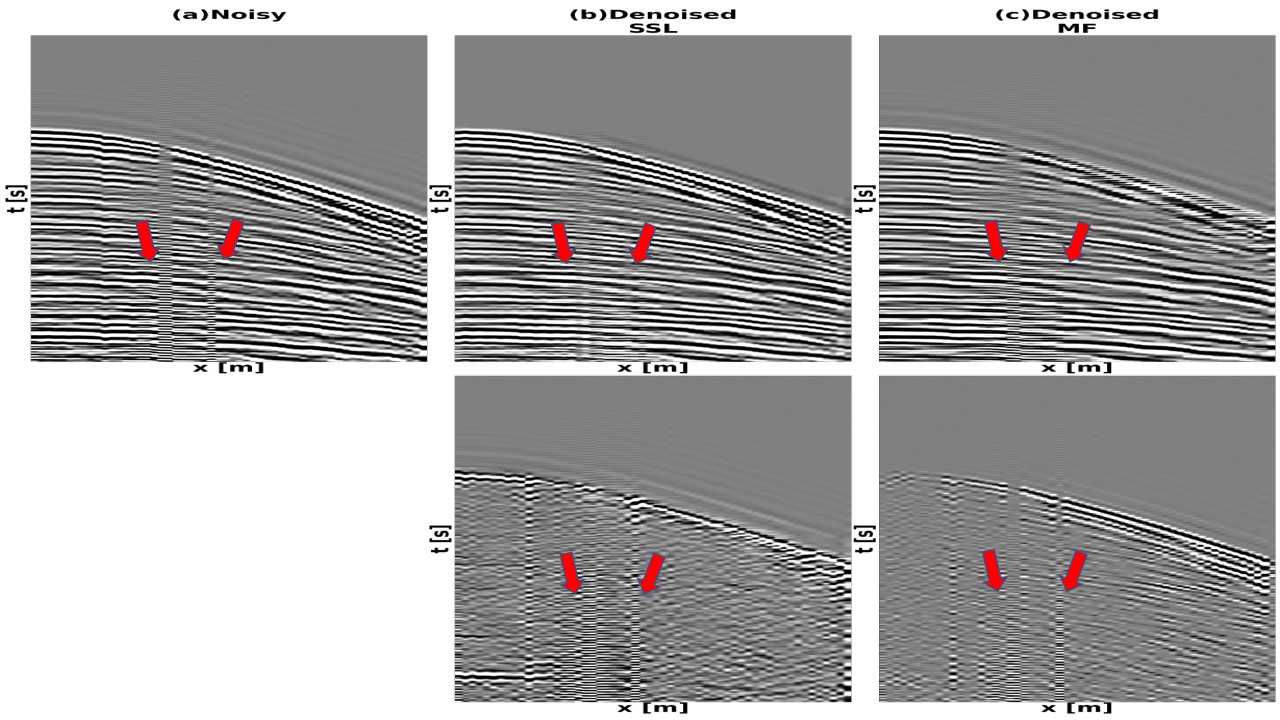}
    \caption{Comparison of the trained SSL model and MF applied to the field dataset with coherent noise.  (a) The  original noisy data, (b) the  result  of  the  SSL  denoising  procedure and difference between the noisy and denoised result, and (c) the  result  of  the  MF and difference between the noisy and denoised result. Red arrows illustrate the original noisy traces in raw data, the corresponding denoised traces and the removed coherent noise, respectively.}
    \label{fig:Comparison}%
\end{figure}

\begin{figure}[htp]
    \centering
    \includegraphics[width=1\textwidth,height=0.8\textwidth]{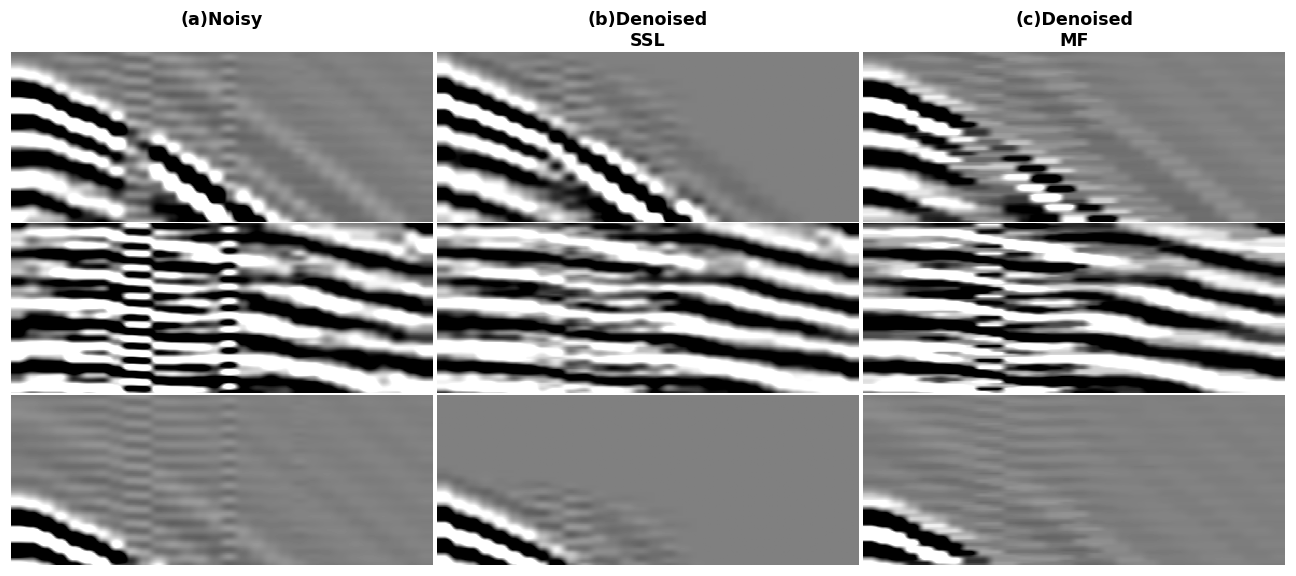}
    \caption{ A close-up of the original data (a), SSL denoised result (b), and MF denoised result (c).}
    \label{fig:close-up}%
\end{figure}

Figure \ref{fig:Comparison}(b) illustrates the result of the trained SSL model applied to the field dataset. For comparative purposes, Figure \ref{fig:Comparison}(c) illustrates the denoising results of a median filter (MF) applied on the raw data, with filter parameters of $3\times3$ footprint. The red arrows in Figure \ref{fig:Comparison} denote the traces 18, 19 and 25 - three noticably noisy traces that exist within the field dataset and that need to be suppressed. Figure \ref{fig:Comparison}(b) and Figure \ref{fig:Comparison}(c) show that the trained SSL network and the MF result in  more consistent reflections across the full shot, than observed in the raw shot. The SSL result is notably more consistent than the MF result in both earlier and later arrivals, indicating that the SSL procedure offers a better denoising performance than the MF. The differences between the original noisy image, the SSL denoised result and the MF result show that the majority of the coherent noise has been removed by both methods with a notably larger noise removal provided by the SSL procedure. The MF causes larger signal leakage at far offsets while the SSL network causes larger signal leakage on later arrivals. For field data, PSNR and SSIM values cannot be computed as we do not have the corresponding noise-free data. Therefore, instead of evaulating the denoised results of field data quantitatively, we judge the result in a zoomed-in comparison for three areas of interest, as depicted in Figure \ref{fig:close-up}. Figure \ref{fig:close-up}(b) shows that the SSL denoised result can fill the trace gaps that are present in the noisy section better and can remove more noise than the MF (Figure \ref{fig:close-up}(c)). Most of the coherent noise has been removed by the two methods, however, both of them  have caused a small amplitude reduction to some of the seismic arrivals.

\vspace{0.00mm}
\section{Discussion}
We have proven blind-trace networks to be efficient suppressors of trace-wise noise in seismic data; and, only requiring noisy images to train the network in a self-supervised way. Adapting the methodology of \cite{broaddus2020removing}, a vertical noise mask is designed in an attempt to suppress trace-wise noise for seismic data.  The network was successfully applied on field data and shows a better denoising performance than a conventional, non-machine learning method-the median filter. As there are some cases that seismic data is severely contaminated by noise, synthetic tests were performed with different noise levels and different numbers of noisy traces. Results show that the network can satisfactorily handle the scenarios of both large noise amplitudes and high numbers of noisy traces in the raw data. The potential of the blind-trace network to suppress other trace-wise noise, such as deblending in seismic data, can be explored based on this paper.

\begin{figure}[htp]
    \centering
    \includegraphics[width=1\textwidth,height=1\textwidth]{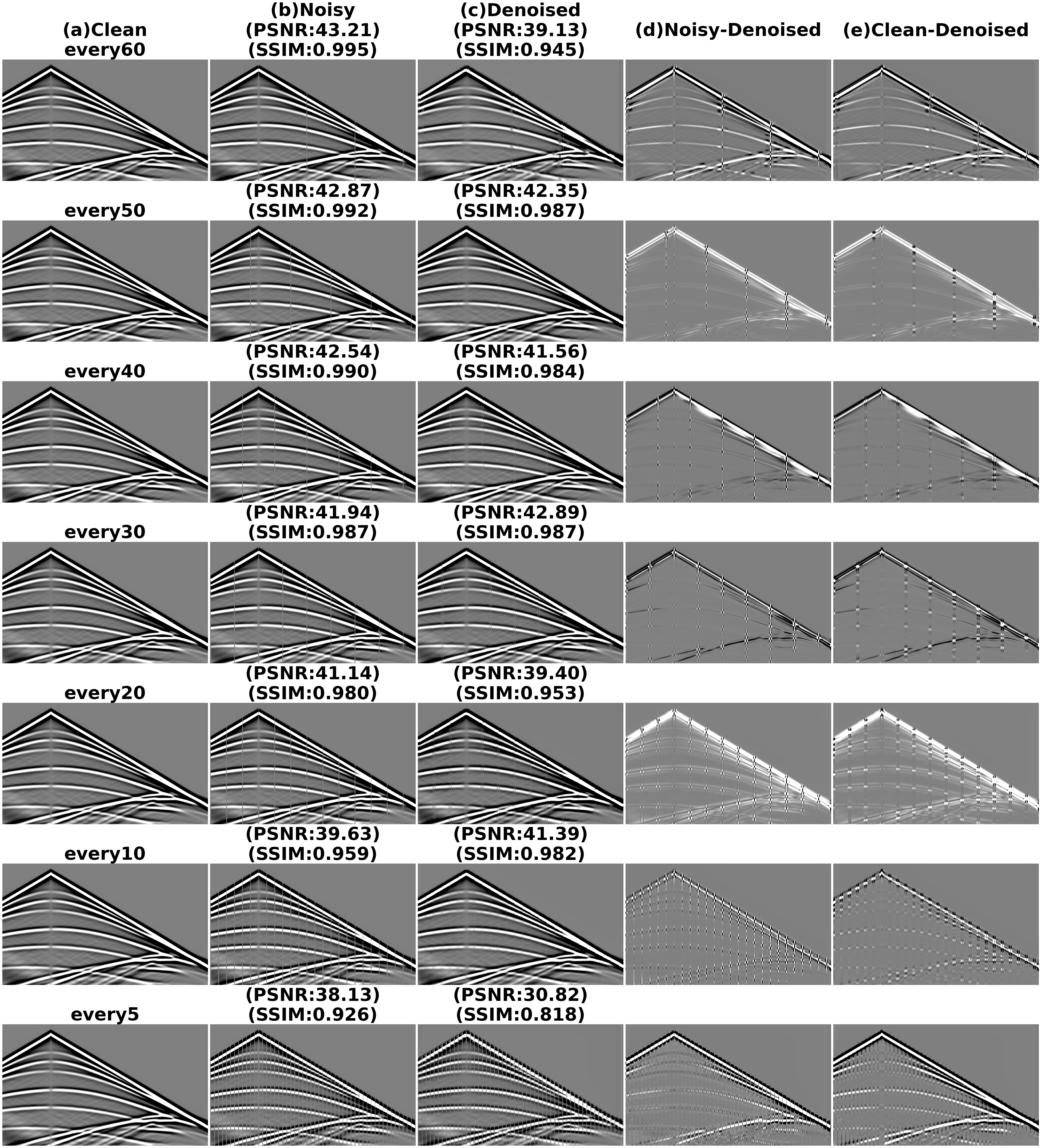}
    \caption{Trained SSL model applied to a synthetic dataset with different patterns of coherent noise.  (a)-(e) The noise-free synthetic, the noisy synthetic given as input to the model, the  result  of  the  SSL  denoising  procedure, the difference between the noisy and denoised datasets and the difference between the noise-free and denoised datasets, respectively, for the seven different noise patterns of contamination. }
    \label{fig:inter}
\end{figure}

Unlike interpolating gathers to recover the missing information within a trace (of value zero) across the receivers, the SSL network with the blind-trace scheme learns to remove the existing trace-wise noise. In this sense, the SSL model is not easily adaptable for regular interpolation tasks due to the computation of the loss being against the original noisy, data. Despite this, we analyse it's ability as a tool for regular interpolation. Similar to the other synthetic experiments, noisy realisations are generated for each shot in the 404 clean patches in Figure 2 by regularly replacing every 60, 50, 40, 30, 20, 10 and 5 traces among the total 256 traces with a constant value of 0 to represent the 404 raw, noisy data as the target (label) for the deep learning network. The test images are noisy realisations generated from the clean data by regularly replacing every 60, 50, 40, 30, 20, 10 and 5 traces with zero values. All the training parameters are as detailed in Table \ref{tab:para} and are the same for the seven different cases. PSNR and SSIM are both used to evaluate the denoising performance, computed against the corresponding noise-free synthetic data. Figure \ref{fig:inter} illustrates the denoised results of different trace-wise noise patterns of contamination for synthetic seismic shot gathers. For all seven cases, the differences between the noisy and denoised datasets demonstrate some amount of noise are removed by the trained network but it struggles to recover the noisy traces completely. The difference between the noise-free and denoised datasets demonstrates that the majority of the noise still remains.

One limitation of the blind scheme is the design of noise mask. In field data, coherent noise often exhibits a relationship in both time and space, which makes it difficult to design a precise shape of noise mask. However, this paper can still provide inspirations to some other different noise types. For example, industrial noise from nearby power stations for carbon dioxide ($\mathrm{CO_2}$) storage applications, as observed by \cite{stork2018assessing}, traffic from nearby roads, as observed by \cite{birnie2016analysis} and \cite{chambers2020using}, and tube waves within borehole monitoring \cite[]{hardage1985vertical}, to mention a few possibilities.
\section{Conclusions}
Blind-trace networks allow a self-supervised
denoising scheme requiring only the noisy seismic data, as opposed to the common requirement of clean data as training labels. Contrary to previous applications where blind-spot networks are shown to be good random denoisers assuming the unpredictability of the targeted noise based on neighbouring values; in this paper, we have proposed an adaptation of the blind-spot strategy to remove temporally-dependent noise. By designing a vertical noise mask, the trace-wise, targeted noise can be suppressed in a self-supervised manner. The proposed method creates training and validation sets based on only the raw noisy shot gathers; to be specific, the raw noisy shot gathers become the network's labels and corrupted versions of the raw shot gathers, created by the blind-trace strategy, become the network's inputs. Such adaptations make the proposed method capable of removing coherent noise through a well-trained CNN model. Experiments on synthetic and field seismic shot gathers have shown that the proposed method can effectively suppress trace-wise coherent noise in seismic data with minimal damage to the signals. Furthermore, we illustrate that the denoising scheme performs well with different levels and different numbers of trace-wise noise contamination, highlighting its robust nature to varying noise levels.
\newpage
\bibliographystyle{plain}  

\bibliography{n2vPaper}
\end{document}